\documentclass{mem}
\usepackage{natbib}
\usepackage{txfonts}
\usepackage{balance}
\usepackage{graphicx}
\usepackage[a4paper,breaklinks,dvipdfm]{hyperref}
\idline{75}{282}
\begin{document}
\def\teff{$T\rm_{eff }$}
\def\kms{$\mathrm {km s}^{-1}$}

\title{
Expect the unexpected: non-equilibrium processes in brown dwarf atmospheres}

   \subtitle{}

\author{
Ch.\,Helling\inst{1}
          }

\institute{
SUPA, University of St Andrews,
North Haugh,
St Andrews, KY16 9SS, UK
\email{ch@leap2010.eu}
}

\authorrunning{Helling}

\titlerunning{Expect the unexpected}

\abstract{Brown Dwarf atmosphere are a chemically extremely rich, one
  example being the formation of clouds driven by the
  phase-non-equilibrium of the atmospheric gas. Cloud formation
  modelling is an integral part of any atmosphere simulation used to
  interpret spectral observations of ultra-cool objects and to
  determine fundamental parameters like log(g) and T$_{\rm eff}$.
  This proceeding to the workshop {\it GAIA and the Unseen: The Brown
    Dwarf Question} first summarizes what a model atmosphere
  simulation is, and then advocates two ideas: A) The use of a
  multitude of model families to determine fundamental parameters with
  realistic confidence interval.  B) To keep an eye on the unexpected,
  like for example, ionisation signatures resulting plasma processes.
  \keywords{Stars: abundances -- Stars: atmospheres} }

\maketitle

\section{Introduction}

\begin{figure*}
{\ }\\*[-0.5cm]
\resizebox{0.9\hsize}{!}{\includegraphics[clip=true, angle=90,origin=c]{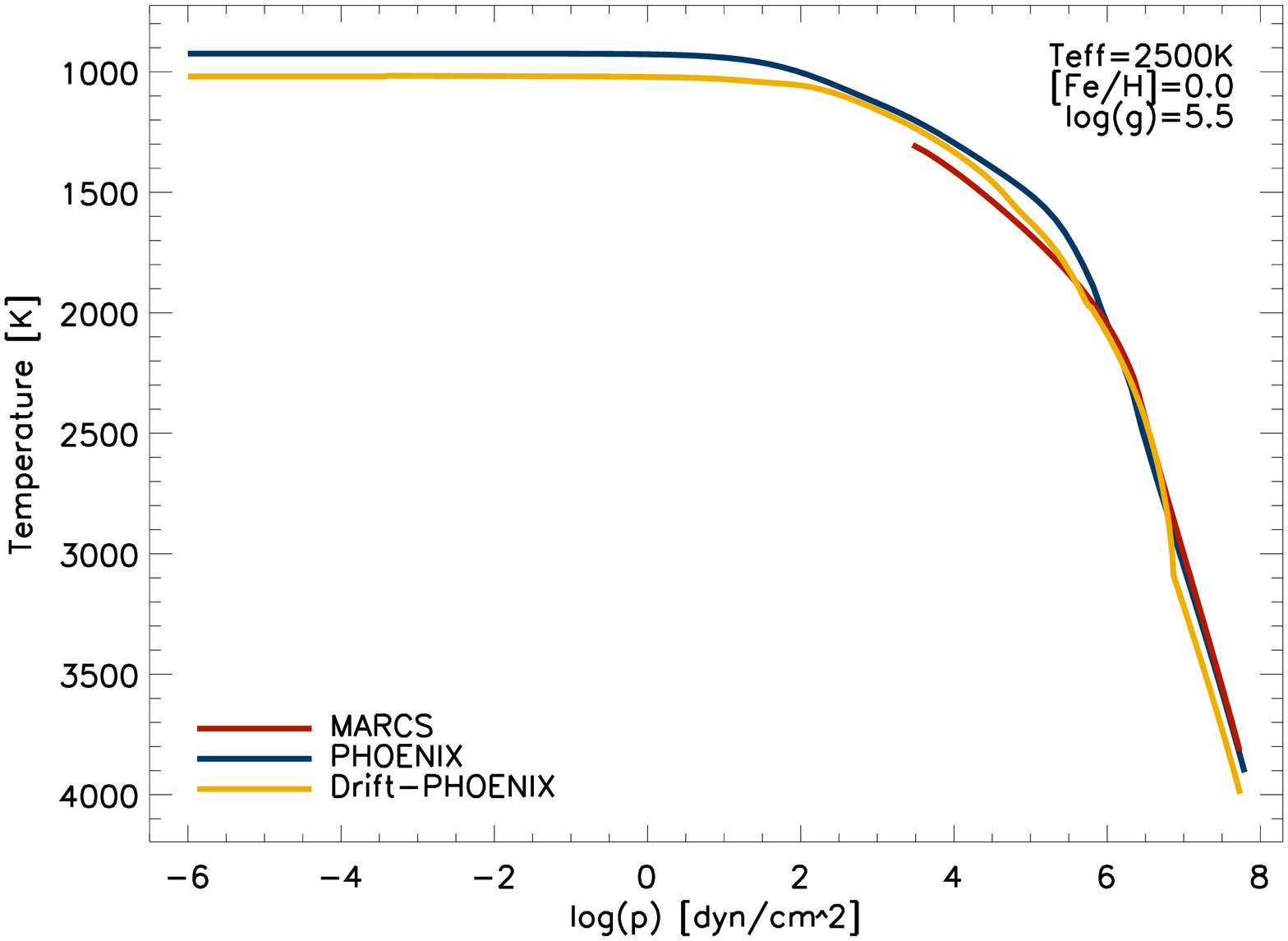}}\\*[-2.3cm]
\resizebox{0.9\hsize}{!}{\includegraphics[clip=true, angle=90,origin=c]{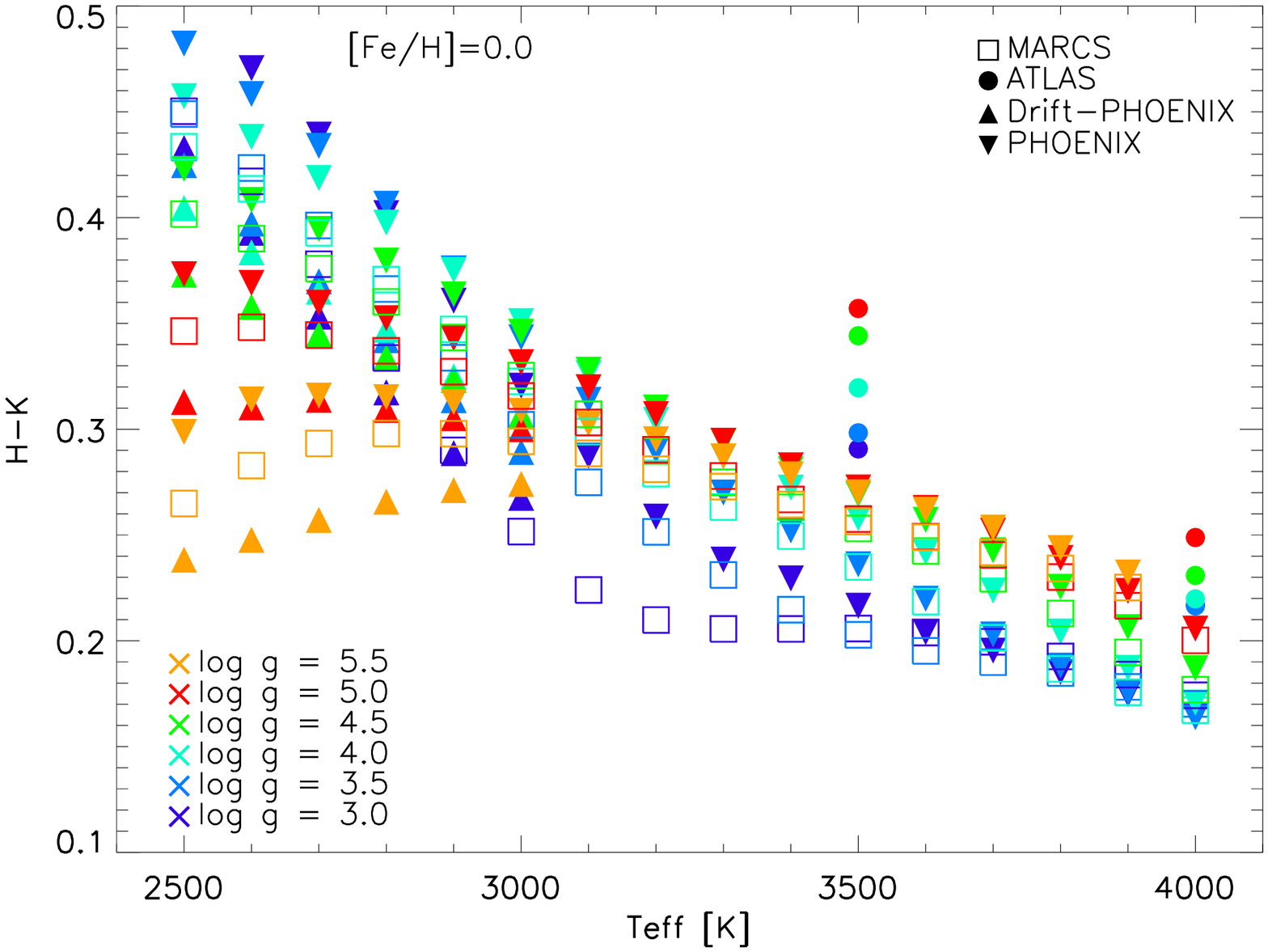}}\\*[-2.2cm]
\caption{\footnotesize {\bf Top:} Differences in the resulting
  (T$_{\rm gas}, p_{\rm gas}$) structure of {\sc MARCS}, {\sc
    Phoenix}, and {\sc Drift-Phoenix} model atmosphere simulations for
  a brown dwarf atmosphere of T$_{\rm eff}$=2500K and log(g)=5.5. {\bf
    Bottom:} Differences in the H-K colour (UKIDSS filters) depending
  on the effective temperature, T$_{\rm eff}$, and surface gravity,
  log(g), for {\sc ATLAS}, {\sc MARCS}, {\sc Phoenix}, and {\sc
    Drift-Phoenix} model atmosphere simulations.\newline The assumed
  element abundances are solar but will differ in detail for the
  individual elements. (Please refer to \cite{boz2014} for detailed
  references to the model atmosphere grids.) }
\label{Tp}
\end{figure*}

Brown dwarfs atmospheres are chemically very active as cloud particles
form inside their atmosphere if the atmospheric gas is in
phase-non-equilibrium. This formation process transforms the
atmosphere into an inhomogeneously depleted gas with an additional
strong opacity component in form of cloud particles. Other processes
that drive the atmosphere out of equilibrium are, for example, the
impact of galactic cosmic rays and rotationally driven winds. Neither
of the latter two is included in any consistent atmosphere model yet

The discovery of the new and unexpected requires to understand the
underlying model assumptions, in this case, the extant to which model
atmosphere simulations are applicable and how to make the most of the
diversity at hand.

Physical modelling and numerical simulations are the backbone of
understanding observational data. Ideally, a consistent description of
physical and chemical processes is aimed for which is determined by a
minimum set of global parameters. Stellar atmosphere modelling has
greatly inspired the brown dwarf modelling community, and hence, the
global parameter normally referred to are the effective temperature,
T$_{\rm eff}$, which represents the total observable flux emitted, the
surface gravity, log(g), the radius or mass, and the element
abundances. These global parameters are linked to the formation (mass,
element abundances) and evolution (T$_{\rm eff}$, element abundances)
of the object. The physical principles at the base of every model
atmosphere are energy, momentum and mass conservation. The solution of
the radiative and convective energy transport provides the local gas
temperature, T$_{\rm gas}$ (e.g. top panel in Fig.~\ref{Tp}), the
convective velocity, v$_{\rm conv}$, for each atmosphere layer and the
wavelength dependent energy distribution F$_{\lambda}$ (the synthetic
spectrum). 1D (brown dwarf) model atmospheres assume hydrostatic
equilibrium which provides the local gas pressure, p$_{\rm gas}$. A
calculation of the chemical composition of the atmosphere for (T$_{\rm
  gas}$(z), p$_{\rm gas}$(z)) (e.g. top panel in Fig.~\ref{Tp}) based
on the pre-scribed element abundances allows the calculation of the
gas-phase opacities that are needed for the radiative transfer
calculation through the atmosphere. The chemical composition, (T$_{\rm
  gas}$(z), p$_{\rm gas}$(z), v$_{\rm conv}$), and the initial element
abundances determine the formation of clouds in brown dwarf and in
planetary atmospheres. Clouds have a strong feedback onto the
atmospheric structure as they deplete element abundances and provide a
strong source for radiative heating and cooling by their opacity.

Several groups do perform such model atmosphere simulations that span
across several spectral types (see references in \citealt{plez2011,
  rojas2013, boz2014}) but comparison studies or the use of different
model atmosphere grid to determine confidence intervals are still
sparse. The use of different model atmosphere grids for data
interpretation should be made good scientific practice in the times of
GAIA and PLATO.  The {\it Virtual Observatory} will soon be providing
the opportunity to apply more than one model atmosphere families to observations
(Taylor 2014). This would also allow an open mind regarding
processes that are not yet included in model atmosphere simulations
(e.g. non-thermal ionisation) and that could help with {\it weather} detections on brown dwarfs (Morales-Calderó et al. 2006, Biller et al. 2014, B\"unzli 2014).

\section{The need of model diversity}

\cite{sarro2013} present a module that will be used to detect and
characterize ultra-cool dwarfs in the {\it Gaia} database. The module
was trained with {\sc Phoenix}-based AMES and BT-settle models for
solar metallicity, and errors suggested range from 10K to 300K. The
energy transfer core module including the gas-phase chemistry is the
same for both model families used. Therefore, the biggest difference
between these {\sc Phoenix} derivatives is the cloud modelling which
is important for T$_{\rm eff}<2700$K. Differences in line list data
will only play a role at high effective temperatures where no clouds
form in the atmosphere.
 
The need for model atmosphere diversity has been demonstrated, for
example, with respect to disk detection. \cite{sincl2010} re-analyzed
far-IR Spitzer data, and they show that the number of disk detections
varies if different model atmospheres were used to
determine the far-IR excess.

\cite{south2012} presents parameters of 38 exoplanets based on an
analysis of homogeneous set of observations. As the author states, the
physical properties of any transiting planet can as yet not be
determined by observing the planet alone but additional constrains are
needed. These constrains are provided through parameters from the host
stars that are determined by applying multiple stellar evolutionary
models (Sect 3.1. in \citealt{south2012}), which then allows to
discuss systematic errors as presented in e.g. Table 4 in
\cite{south2012} (see also the paper's Appendix).

The work by \cite{boz2014} suggest that the difference in model
atmosphere results can be used to provide a better estimate of the
confidence interval for planetary equilibrium temperatures and for the
location of the habitable zone around M dwarfs. Both measures are
related to the sustainability of life-important chemical species. A
change of only ~20K can already hinder the existence of liquid water
on a planetary surface. We note, that it is very unrealistic to claim
that any of the model atmospheres can achieve such a accuracy in
predicting global parameters with such a precision from observed
spectra. Comparative spectrum fitting as in
\cite{dup2010,pat2012,bon2014} demonstrate this clearly.
\cite{kane2014} demonstrates the uncertainty of the habitable zones location
resulting from stellar parameter uncertainties for confirmed
exoplanetary host stars and {\it Kepler} candidate hosts.

Figure~\ref{Tp} (top) shows as an example the comparison of the local gas
temperature - gas pressure profile (T$_{\rm gas}, p_{\rm gas}$) for a
brown dwarf atmosphere simulation from three different model
families. Each of these three models represents one symbol in
Fig.~\ref{Tp} (bottom)  which shows the differences between the model families
in H-K colour plotted for the available effective temperatures,
T$_{\rm eff}$, and surface gravities, log(g). Differences are largest
for low T$_{\rm eff}$, but this reflects more the differences between
cloud-free (ATLAS, MARCS, {\sc Phoenix}) and cloudy models ({\sc
  Drift-Phoenix}). The differences at the high-temperature end of the
T$_{\rm eff}$-axis are suggested to result from differences in
gas-opacity data, element and chemical abundances, convective
treatment.

\cite{plav2014} note that such model differences may be responsible
for uncertainties of planetary radii as derived for planets in the
{\it Kepler} sample. They suggest that improving stellar parameters is
essential for resolving this radius issue, which hence requires a
sensible use of model atmospheres results.

\section{The unexpected}

Brown dwarfs seemed the perfect example for a static atmosphere, until
it was understood that cloud formation plays a major role for their
atmospheres (\citealt{tsuji1996}). Then, older brown dwarfs were
though to be the perfect example for a neutral atmosphere, until it
was suggested that detections of radio emission (\citealt{berg2001})
should be related to ionisation processes inside the atmosphere
(\citealt{hell2011a}). Meanwhile, different processes were shown to
contribute to the increase of the local ionisation, including
wind-driven gas ionisation (\citealt{stark2013}) and gas and cloud
particle ionisation by cosmic rays (\citealt{rimmer2013}). The
atmospheric clouds can easily ionise in a turbulent atmosphere
(\citealt{hell2011b}) leading to electron or dust dominated discharge
regimes (Fig. 10 in \citealt{hell2013}). If a small-scale discharge
successfully sets of, an ionisation front develops that travels
through the atmospheric gas and eventually emerges as a large-scale
lightning or sprite. \cite{bailey2014} present a table of typical
lightning signatures and refer to detectors for possible observations
on Earth. This suggest that similar observations might be possible for
brown dwarfs in the future, or that such signatures may be present but
hidden in existing data.

\section{Conclusions}
As a result of the Gaia workshop, {\sc Drifit-Phoenix} will be included
into the model atmosphere database of the Virtual Observatory to allow
a multi-model approach to observations.

\begin{acknowledgements}
Amelia Bayo is thanked for her initiative to make {\sc Drfit-Phoenix}
synthetic spectra part of the {\it Virtual Observatory}. ChH
highlights financial support of the European Community under the FP7
by an ERC starting grant 257341.  Most literature search was performed using
the ADS. Our local computer support is acknowledged highly.
\end{acknowledgements}

\bibliographystyle{aa}

\end{document}